\documentclass[useAMS,usenatbib]{mn2e}
\usepackage[dvips]{graphicx}
\usepackage{longtable}

\title[Multiband Optical Photometry of SN 2004S]{Multiband Optical Photometry 
and Bolometric Light Curve of the Type Ia Supernova 2004S }
 
\author[Kuntal Misra et al.]
{Kuntal Misra$^1$, Atish P. Kamble$^2$, D. Bhattacharya$^2$ and Ram Sagar$^1$\\
1.Aryabhatta Research Institute of Observational Sciences, Manora Peak, Nainital, 263 129, India\\
2.Raman Research Institute, Bangalore, 560 080, India\\
(E-mail:kuntal{@}upso.ernet.in, atish{@}rri.res.in, dipankar{@}rri.res.in,
sagar{@}upso.ernet.in)}
 
\begin{document}

\date{Accepted.....; Received .....}

\pagerange{\pageref{firstpage}--\pageref{lastpage}} \pubyear{}

\maketitle

\label{firstpage}

\begin{abstract}
 We present {$\rm B$}{$\rm V$}{$\rm R_{c}$}{$\rm I_{c}$} broad band
CCD photometry of the Type Ia supernova SN 2004S, which appeared in the 
galaxy MCG-05-16-021, obtained during 2004 February 12 to March 22. 
Multiband and bolometric light curves constructed using our data as well 
as other available data are presented. The time of B band maximum and the 
peak magnitudes in different bands are obtained using the fits of light 
curve and colour templates. We clearly see a strong shoulder in $\rm R_{c}$
 band and a second maximum in $\rm I_{c}$ band. SN 2004S closely 
resembles SN 1992al after maximum. From the peak bolometric luminosity we 
estimate the ejected mass of $\rm ^{\rm 56}Ni$ to be 0.41 $\rm M_{\odot}$.

\end{abstract}
 
 
\section{INTRODUCTION}

In recent years significant progress has been made in the study of Type Ia 
Supernovae (SNe), but many of their properties remain fairly uncertain. 
Supernovae of Type Ia are among the most luminous stellar outbursts and 
because of the homogeneity in their properties \citep{Hoflich}  have been 
regarded as standardizable candles for determining extragalactic distances and 
deriving cosmological parameters. They are thought to be thermonuclear 
explosions of carbon-oxygen white dwarfs \citep{HF}. However, Type Ia 
supernovae are suspected to be not a perfectly homogeneous group, from both 
their light curves and spectra \citep{Pskovskii1, Pskovskii2,
Barbon1, Barbon2, Branch1, Elias2, Frogel, Phillips1, Cristiani}.
Some SNe have shown significant deviations such as SN 1991T 
\citep{Filippenko1, Phillips2, Jeffery, Mazzali1} and
SN 1991bg \citep{Filippenko2, Leibundgut2,Turatto, Mazzali2}.
 The classic Si II and Ca II lines were seen very late and with
 diminished strength in SN 1991T 
and its early spectrum was dominated by Fe III lines
\citep{Filippenko1, Ruiz-Lapuente} while the nebular phase
was very similar to other SN Ia \citep{Leibundgut2}. SN 1991bg was a strongly 
subluminous event which established the existence of a wide range of 
luminosities among Type Ia supernovae \citep{Filippenko2, Leibundgut2}.
SN 1991bg showed an absorption trough near 4000\AA which was attributed to 
Ti II ($\lambda$$\lambda$ 4395\AA, 4444\AA and 4468\AA) absorption 
\citep{Filippenko2, Mazzali2}. Other supernovae showing remarkable deviations 
are SN 1999ac, a slow rise and fast decliner \citep{Labbe, Phillips5}, 
SN 2000cx, a fast riser and slow decliner had unusually blue (B-V) colours 
$\sim$ 30 days after blue maximum \citep{Li1, Candia},
while SN 1986G \citep{Phillips1} appeared to have properties between normal 
supernova and the extreme case of SN 1991bg. 
SN 1999by is a rare example of a ``peculiar", fast declining SN Ia. Recently 
\citet{Garnavich} presented detailed photometric and spectroscopic observations 
of SN 1999by. It is one of the few SNe to show significant intrinsic 
polarization \citep{Howell}. \citet{Li3} describe the even stranger SN 2002cx,
 which had premaximum spectra like 1991T, a luminosity like SN 1991bg 
(subluminous event), a slow late time decline and unidentified spectral lines. 
In spite of these differences in SNe Ia, they still seem to follow a few 
common patterns in their behavior \citep{Leibundgut3}. Of these, the 
correlation between the linear decline rate and luminosity is the best known 
\citep{Phillips3}. 
The template fitting or $\Delta${$m_{15}$} (the number of magnitudes in B band 
by which the SN declines in the first 15 days after maximum) method 
\citep{Hamuy1, Phillips4}, the multi-colour light curve shape correction 
\citep{Riess1,Riess2}, and the stretch factor \citep{Perlmutter} 
exploit this property of SNe Ia to determine their luminosities.

In this paper, we report optical photometry of the Type Ia supernova 
SN 2004S. This supernova (mag 16 on red CCD images) was discovered on 2004 
February 3.54 UT by \citet{Martin} at Perth Observatory with the 0.61 - m 
Perth/Lowell Automated telescope in the course of the Perth Automated 
Supernova Search. The position of the supernova was reported by \citet{Biggs}
 to be R.A. = $06^{\rm h}$$45^{\rm m}$$43^{\rm s}$.50 $\pm$ $0^{''}.1$, 
Dec. = $-31^{\circ}13^{'}52^{''}.5~ \pm$ $0^{''}.1$ (J2000).  
The SN is situated $47^{''}.2$ W and $2^{''}.5$ S of the galaxy  
MCG-05-16-021. Spectrophotometry obtained at CTIO by \citet{Suntzeff2}
on February 6.1 UT identifies it as a type Ia supernova with an 
expansion velocity of $\sim$ 9300 km/sec.

We have carried out multi-colour optical photometric observations during 
the early decline phase. We have used these in combination with data available 
in the literature to study the development of the optical light curve.
 The details of the 
photometric observations are presented in the next section while the 
development of the light curve and other properties of the supernova are 
discussed in the sections to follow.

\section{OBSERVATIONS AND DATA REDUCTION}
 \begin{figure}
\includegraphics[width=84mm]{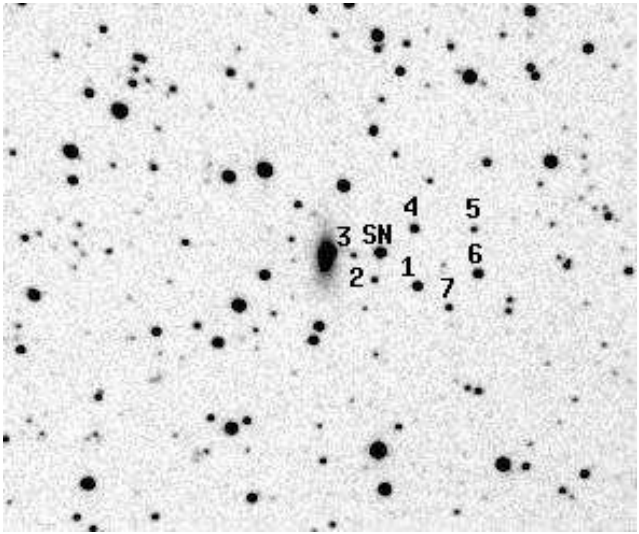}
\caption{SN 2004S and the comparison stars.}
\label{figure1}
\end{figure}

 We began optical photometry of SN 2004S approximately eight days after 
the discovery. The observations were carried out from ARIES, Nainital, India 
at 27 epochs during the period 2004 February 12 to March 22 using a 
1024 $\times$ 1024 $\rm pixel^{2}$ CCD camera attached to the f/13 Cassegrain 
focus of the 104-cm Sampurnanand Telescope. One pixel of the CCD chip 
corresponds to a square of $\sim$ 0.38 arcsec while the entire chip covers a 
field of 6 $\times$ 6 $\rm arcmin^{2}$ on the sky. The gain and read out noise
 of the CCD camera are 12 electrons per Analogue to Digital Unit (ADU) and 
7 electrons respectively. Due to its large negative declination the supernova 
had to be observed at a low elevation through large airmass. This precluded U 
band observations, but we were able to carry out photometry at 
BV$\rm R_{c}$$\rm I_{c}$ bands during most of the nights until March 22, 
beyond which it became too faint to follow. Exposure times in the early decline
 phase of the B,V, ${\rm R_{c}}$ and ${\rm I_{c}}$ band images were 300, 300, 
120 and 120s respectively whereas the exposure time was increased considerably
 in the later phase so as to get a good signal. The exposure time in this phase 
varied from 10 min in ${\rm R_{c}}$ and ${\rm I_{c}}$  to 30 min in B and 
V bands. We observed several bias and twilight flat frames with the CCD camera 
to calibrate the supernova images using standard techniques. Data reduction was
 carried out using IRAF\footnote{IRAF is distributed by the National Optical 
Astronomy Observatory, USA.} and MIDAS softwares. The images were bias 
corrected and flat fielded using the CCD reduction package in IRAF.
For photometric calibrations we have used comparison stars 1, 2, 4, 5, 6 and 7 
of \citet{Krisciunas1}. The standard magnitudes for the local calibrators as 
determined by \citet{Krisciunas1} are given in Table 1 corresponding to the 
stars marked in the finder chart in Figure 1. The calibration stars were 
observed along with SN 2004S. The instrumental magnitudes of SN 2004S and the 
comparison stars were estimated using aperture photometry in IRAF. We 
determined the zeropoints for our system on each observing 
night by matching the instrumental magnitudes to the values given in Table 1 
to obtain the standard magnitude of the SN. The difference in the 
measured ($\rm B_{obs}$, $\rm V_{obs}$, $\rm R_{{c}_{obs}}$ and 
$\rm I_{{c}_{obs}}$)  and standard ($\rm B_{st}$, $\rm V_{st}$, 
$\rm R_{{c}_{st}}$ and $\rm I_{{c}_{st}}$) BV${\rm R_{c}}$${\rm I_{c}}$ 
magnitudes of calibration star 1 is plotted in Figure 2. We see that the 
difference is consistent with zero, with standard deviations 0.02, 0.04, 
0.05 and 0.06 mag in B, V, ${\rm R_{c}}$ and ${\rm I_{c}}$ passbands 
respectively. These standard deviations were added in quadrature to the 
instrumental errors in order to get the final estimates of error in the 
determined magnitudes in different bands. The resulting 
BV${\rm R_{c}}$${\rm I_{c}}$ magnitudes of SN 2004S based on our observations 
are provided in Table 2 along with errors.

\begin{table}
\caption{Adopted BVRI magnitudes of comparison stars. Star Numbers correspond 
to those marked in figure 1.}
\noindent
\begin{tabular}{@{}ccccc}\hline
Star No.& B & V & R & I\\
\hline
1&16.44 &15.61 &15.16 &14.74\\
2&18.50 &17.45 &16.79 &16.22\\
3&18.18 &17.60 &17.30 &16.91\\
4&17.89 &16.53 &15.65 &14.86\\
5&18.55 &17.52 &16.88 &16.33\\
6&16.28 &15.59 &15.19 &14.78\\
7&17.85 &17.00 &16.51 &16.07\\
\hline
\end{tabular}
\label{table 1}
\end{table}

 \begin{figure}
\includegraphics[width=84mm,height=84mm]{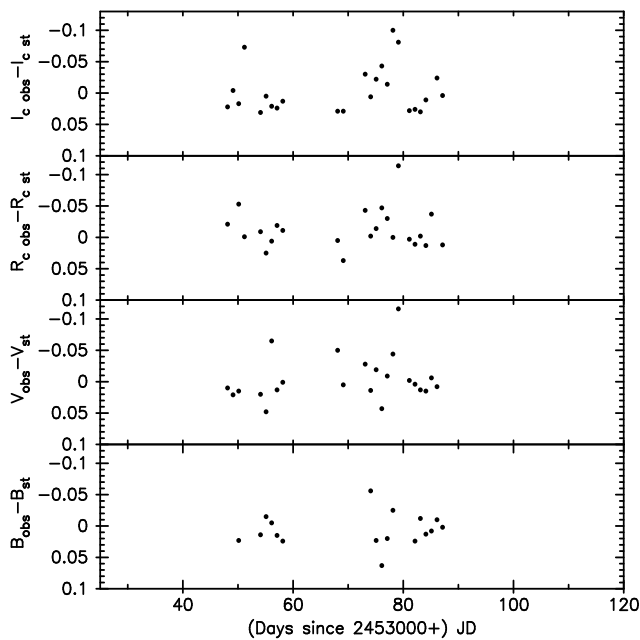}
\caption{Scatter in the estimated magnitude of calibration star 1 during multiple observing nights.}
\label{figure2}
\end{figure}
\begin{table*}
\caption{BV${\rm R_{c}}$ and ${\rm I_{c}}$ magnitudes of SN 2004S along with 
errors, Julian date and mid UT of observations are listed.}
\noindent
\begin{tabular}{cccccc}\hline
Date (UT)& Time in JD&B (mag)&V (mag)&$\rm R_{c}$ (mag)& $\rm I_{c}$ (mag)\\
\hline
&& & & & \\
2004 02 12.614& 2453048.1141& 14.85 $\pm$ 0.022& 14.58 $\pm$ 0.041& 14.52 $\pm$ 0.051& 14.76 $\pm$ 0.062\\
2004 02 13.608& 2453049.1088&         $-$      & 14.75 $\pm$ 0.043&       $-$        & 14.88 $\pm$ 0.061\\
2004 02 14.632& 2453050.1328& 15.17 $\pm$ 0.022& 14.75 $\pm$ 0.042& 14.64 $\pm$ 0.051& 14.85 $\pm$ 0.063\\
2004 02 15.677& 2453051.1773&        $-$       &       $-$        & 14.77 $\pm$ 0.052& 14.85 $\pm$ 0.061\\
2004 02 18.601& 2453054.1011& 15.48 $\pm$ 0.022& 14.91 $\pm$ 0.041& 14.91 $\pm$ 0.050& 15.12 $\pm$ 0.061\\
2004 02 19.605& 2453055.1055& 15.71 $\pm$ 0.023& 14.98 $\pm$ 0.040& 15.04 $\pm$ 0.051& 15.11 $\pm$ 0.061\\
2004 02 20.607& 2453056.1070& 15.76 $\pm$ 0.022& 15.01 $\pm$ 0.040& 15.01 $\pm$ 0.050& 15.14 $\pm$ 0.061\\
2004 02 21.592& 2453057.0920& 15.88 $\pm$ 0.022& 15.07 $\pm$ 0.040& 15.00 $\pm$ 0.052& 15.15 $\pm$ 0.061\\
2004 02 22.632& 2453058.1327& 16.00 $\pm$ 0.022& 15.13 $\pm$ 0.040& 15.07 $\pm$ 0.050& 15.10 $\pm$ 0.061\\
2004 03 01.618& 2453066.1185&        $-$       & 15.53 $\pm$ 0.042& 15.15 $\pm$ 0.051&       $-$ \\
2004 03 03.587& 2453068.0885&        $-$       & 15.56 $\pm$ 0.041& 15.20 $\pm$ 0.050& 14.96 $\pm$ 0.061\\
2004 03 04.612& 2453069.1134&        $-$       & 15.78 $\pm$ 0.042& 15.37 $\pm$ 0.051& 15.08 $\pm$ 0.061\\
2004 03 08.591& 2453073.0913&        $-$       & 15.99 $\pm$ 0.049& 15.56 $\pm$ 0.066& 15.18 $\pm$ 0.067\\
2004 03 09.580& 2453074.0823& 17.23 $\pm$ 0.033& 16.02 $\pm$ 0.041& 15.62 $\pm$ 0.051& 15.28 $\pm$ 0.061\\
2004 03 10.579& 2453075.0814& 17.31 $\pm$ 0.026& 16.11 $\pm$ 0.041& 15.73 $\pm$ 0.051& 15.34 $\pm$ 0.062\\
2004 03 11.600& 2453076.1041& 17.32 $\pm$ 0.028& 16.13 $\pm$ 0.042& 15.81 $\pm$ 0.052& 15.38 $\pm$ 0.064\\
2004 03 12.597& 2453077.1011& 17.41 $\pm$ 0.030& 16.20 $\pm$ 0.041& 15.87 $\pm$ 0.051& 15.48 $\pm$ 0.061\\
2004 03 13.611& 2453078.1154& 17.47 $\pm$ 0.029& 16.25 $\pm$ 0.041& 15.89 $\pm$ 0.051& 15.47 $\pm$ 0.061\\
2004 03 14.614& 2453079.1182& 17.58 $\pm$ 0.028& 16.30 $\pm$ 0.042& 15.94 $\pm$ 0.051& 15.63 $\pm$ 0.062\\
2004 03 16.602& 2453081.1069& 17.59 $\pm$ 0.038& 16.41 $\pm$ 0.044& 16.05 $\pm$ 0.052& 15.74 $\pm$ 0.062\\
2004 03 17.615& 2453082.1200& 17.62 $\pm$ 0.028& 16.43 $\pm$ 0.041& 16.11 $\pm$ 0.051& 15.80 $\pm$ 0.061\\
2004 03 18.598& 2453083.1024& 17.63 $\pm$ 0.027& 16.43 $\pm$ 0.041& 16.13 $\pm$ 0.051& 15.83 $\pm$ 0.061\\
2004 03 19.594& 2453084.0988& 17.66 $\pm$ 0.030& 16.51 $\pm$ 0.041& 16.17 $\pm$ 0.051& 15.88 $\pm$ 0.062\\
2004 03 20.597& 2453085.1018& 17.73 $\pm$ 0.031& 16.52 $\pm$ 0.043& 16.23 $\pm$ 0.052& 15.96 $\pm$ 0.062\\
2004 03 21.614& 2453086.1184& 17.74 $\pm$ 0.028& 16.56 $\pm$ 0.041& 16.25 $\pm$ 0.051& 15.99 $\pm$ 0.061\\
2004 03 22.615& 2453087.1193& 17.74 $\pm$ 0.030& 16.56 $\pm$ 0.042& 16.30 $\pm$ 0.051& 16.05 $\pm$ 0.062\\
&& & & &\\

\hline
\end{tabular}
\label{table 2}
\end{table*}

\section{UBV$\rm {R_{c}}$${\rm I_{c}}$ light curves and Color Curves }

Our observations started several days after the discovery so we do not have 
observations near peak brightness. To estimate the peak magnitude 
and the peak time in different bands by making template fits it is important 
to have observations temporally as close to the peak as possible. Late 
time observations are also important to perform template fitting of the SN 
light curves. For this purpose, we have used observations of SN 2004S reported 
elsewhere. This also allows us to cross compare our photometry with other 
available data in the literature. The U band observations taken from
 the literature help us 
determine the total luminosity at selected epochs and hence construct the 
bolometric light curve. Our observations present a temporally dense coverage. 
The frequency distribution of our data is 
N(B, V, ${\rm R_{c}}$, ${\rm I_{c}}$) = (20, 25, 25, 25). The other 
observations have been taken from compilations at
http://www.astrosurf.com/snweb2/2004/04S\_/04S\_Meas.htm with contributions 
from \citet{Krisciunas2}, \citet{Espinoza}, \citet{Santallo} and \citet{Lacruz}.
The frequency distribution of the data taken from the literature is 
N(U, B, V, ${\rm R_{c}}$, ${\rm I_{c}}$) = (18, 20, 20, 23, 20).  
The BV${\rm R_{c}}$${\rm I_{c}}$ light curves are shown in Figure 3. U band 
data taken from the literature are also included for comparison. 
To determine the value of $\chi^{2}$ in template fitting we have assumed an 
error of 0.05 mag in all the data points available in the literature.

Since we do not have observations around peak brightness we adopted the template
fitting method to get the magnitudes at peak. We attempted to fit the different
 template sets given by \citet{Riess1} and \citet{Hamuy2}. 
\citet{Hamuy2} present a family of six BVI templates produced from CCD 
photometry of seven well-observed events (1992bc, 1991T, 1992al, 1992A, 1992bo,
1993H and 1991bg). These templates were fit to our observed data using a
$\chi$$^{2}$ minimizing technique which solved simultaneously for the epoch of 
maximum brightness in blue band $\rm t_{B_{max}}$ and the magnitudes 
$\rm B(t_{B_{max}})$, $\rm V(t_{B_{max}})$ and $\rm I_{c}(t_{B_{max}})$. 
We refer to these magnitudes as $\rm B(t_{B_{max}})$, $\rm V(t_{B_{max}})$
 and $\rm I_{c}(t_{B_{max}})$ in the paper. One set of BVI templates from
 \citet{Hamuy2}, that for SN 1992al,  provided a much better fit, as judged
 by the value of the reduced $\chi$$^{2}$, than all others. We also found that
 the SN 1992al template \citep{Hamuy2} fit our data better than the 
parametrized multiband templates given by \citet{Riess1}. The time for 
$\rm B_{max}$ obtained using templates given by \citet{Riess1} was two days 
later than that obtained from \citet{Hamuy2} templates. The similarity in 
behavior between SN 2004S and SN 1992al is remarkable. Even in $\rm I_{c}$ band,
 where the difference between SN Ia are more pronounced \citep{Suntzeff1}, 
SN 2004S follows SN 1992al very closely.

Since the complete data set includes data from different telescopes
and different filter systems in different instruments there can be
systematic differences in the estimated magnitudes \citep{Suntzeff3}.
For example,
\citet{Stritzinger} find systematic difference of ~ 0.05 mag
in photometry by two different telescopes (CTIO and YALO), even
though the photometry is reduced to the same local standards.
A method called ``S-correction",
to bring photometry to a standard system, in such cases was suggested by
\citet{Stritzinger}. \citet{Krisciunas3} and
\citet{Krisciunas4} have applied these corrections to
photometry of various SNe obtained using CTIO 0.9m and YALO
telescopes. To assess any such systematic differences between our data
set and that from literature, we made B, V, I template fits,
discussed above, to these data sets independently.
We found no systematic difference in B band magnitudes of the two sets.
However, our V magnitudes are fainter by 0.03 mag and $\rm I_{c}$ magnitudes are
brighter by 0.05 mag compared to the literature data set. These differences are
comparable to our observational errors. Hence, we do not find it
necessary to apply S - correction to our data set while fitting
templates to the combined data.

In Figure 3 we have included the BV$\rm I_{c}$ template fits to the data. The 
main parameters of SN 2004S as estimated from template fits are listed in 
Table 4. For a typical SN Ia a two day difference is seen between the times 
of B and V maximum \citep{Leibundgut1}. According to the best fit template, 
SN 2004S would have reached maximum brightness in $\rm I_{c}$ band slightly 
earlier than in B band and roughly two days later in V band.

Since our observations started $\sim 8$ days after the peak in the B band,
we do not have observations at or around the epoch of B band maximum.
An excellent match of the SN 1992al light curve shape with that of SN 2004S
indicates that the peak in B band occurred at JD 2453039.42. Though individual 
SNe can be different in their light curve shapes, it seems unlikely that 
the peak magnitudes $\rm B(t_{B_{max}})$, $\rm V(t_{B_{max}})$ and 
$\rm I_{c}(t_{B_{max}})$ of SN 2004S as inferred from the overall match of 
light curves with those of SN 1992al would be much in error. As a consistency 
check, we compare the colours of SN 2004S at this epoch with those obtained using
the intrinsic colour curves of SN Ia population given by \citet{Nobili}. 

\citet{Nobili} present the intrinsic colour curves for a sample of 48 well 
observed nearby SN Ia for 40 days from the epoch of $\rm B_{max}$. We estimate
total selective extinction along the line of sight towards SN 2004S comparing 
the observed colours with the intrinsic colour curves given by \citet{Nobili}. 
Corresponding selective extinctions were taken as fit parameters.
Best fit values of total selective extinctions thus obtained are
listed in Column 2 of Table 3. The shapes of observed colour curves are similar 
to the intrinsic colour curves given by \citet{Nobili} except for systematic 
shifts in individual colour curves due to selective extinction. In Figure 4, 
observed colour curves are plotted over intrinsic colour curves, corrected for 
the best fit values of selective extinction. 

Independently, light curve template fitting gives $\rm B(t_{B_{max}})$, 
$\rm V(t_{B_{max}})$ and $\rm I_{c}(t_{B_{max}})$. From these we calculate 
another set of $\rm E(B-V), E(B-I_{c}), E(V-I_{c})$ using intrinsic colours at 
$\rm t_{B_{max}}$ \citep{Nobili}. These values are listed in Column 3 of 
Table 3. Comparison of the two sets of selective extinction values shows that
the observed colours at $\rm t_{B_{max}}$ are consistent with the intrinsic 
colours given by \citet{Nobili}. We find that these values are also
consistent with the fitted values in Column 2 of Table 3. 

We next calculate the amount of selective extinction expected from galactic 
extinction law, all along the line of sight towards SN 2004S. 
Using $\rm R_{v}$ = 3.1 and $\rm E(B-V)$ = 0.18 $\pm$ 0.054 mag, as obtained from 
fits to colour curves, values of selective extinctions in other colours are 
calculated and listed in Column 4 of Table 3. We find that these values are 
consistent, within errors, with those obtained from the colour curve fits listed 
as Column 2 of Table 3. The estimated reddening in this direction due to our own
 galaxy from \citet{Schlegel} is E(B-V) = 0.101 mag. So a small amount of 
extinction could arise in the host galaxy of SN 2004S.

Thus, we conclude that the magnitudes at $\rm t_{B_{max}}$ as obtained using 
the templates of SN 1992al are fairly representative of the sample SN Ia of 
\citet{Hamuy2}. Further, we use these magnitudes to calculate the peak 
luminosity of SN 2004S in $\S$ 4.

\citet{Hamuy2} do not present $\rm R_{c}$ band templates. In order to compare 
our $\rm R_{c}$ band light curve we construct an expected light curve using 
$\rm (V-R_{c})$ intrinsic colours \citep{Nobili} and the V band template of 
SN 1992al \citep{Hamuy2}. We use for selective extinction, $\rm E(V-R_{c})$ 
listed in Column 2 of Table 3. This derived light curve is plotted in Figure 3.

As seen in Figure 3, except for a shoulder $\sim 26$ days after 
$\rm B({t_{B_{max}}})$, the derived $\rm R_{c}$ band light curve represents 
the observed data well.

Also seen in Figure 3, is a pronounced second maximum in 
$\rm {I_{c}}$ band displayed by SN 2004S. This second maximum is reached 
nearly 26 days after the B maximum. The magnitude of the second 
$\rm {I_{c}}$ maximum is given in Table 4. Such behavior has also been noted 
for some other SNe Ia \citep{Ford, Suntzeff1,Lira, Meikle, Elias1, Elias2,Li1}. 
The second maxima in the $\rm {I_{c}}$ band light curves are usually attributed
 to a temporary increase in absorption which reduces with the fall in the 
degree of ionization several weeks after maximum light \citep{Elias1, Pinto}.

\begin{figure}
\includegraphics[width=84mm,height=84mm,angle=-90]{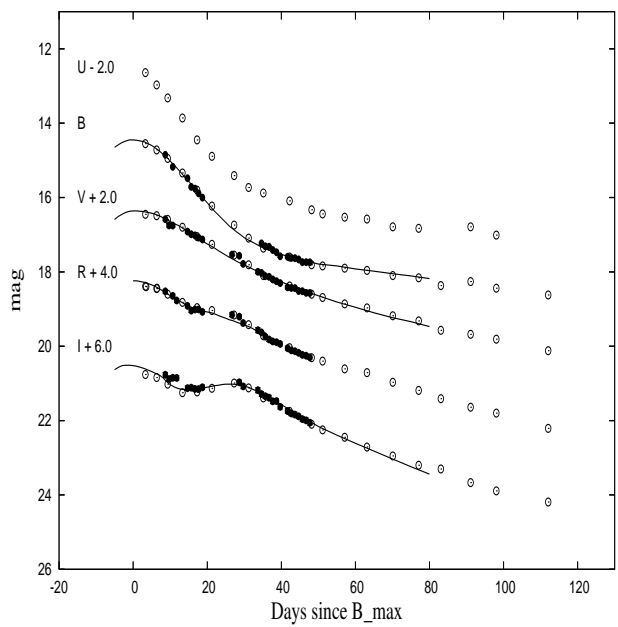}
\caption{UBV$\rm R_{c}$$\rm I_{c}$ light curve of SN 2004S. The light curves 
are offset by a constant value on the magnitude scale as indicated in the plot.
 Filled circles represent our data and open circles represent data from
\citet{Krisciunas2}. Uncertainties in the data are smaller than the size of 
the points.}
\label{figure3}
\end{figure}

\begin{table}
\caption{Selective extinction along the line of sight towards SN 2004S. 
	Column 2 are the values obtained as a fit to the observed and intrinsic
 colours.
	Column 3 are the values at $\rm t(B_{max})$ obtained from the best fit 
template, intrinsic colours and galactic extinction law $\rm R_{v} = 3.1$.
Column 4 are the values obtained using best fit E(B$-$V) from colour 
curves and galactic extinction law $\rm R_{v} = 3.1$.}
\begin{tabular}{@{}llll}
\hline
			& from fits to 		& at $\rm t(B_{max})$ :	& galactic  \\
			& colour curves		& template fits		& extinction law	\\
\hline
E(B$-$V) 	 	& 0.18 $\pm$ 0.054	& 0.20  $\pm$ 0.095	& 0.18  $\pm$ 0.054\\
E(B$-\rm I_{c}$) 	& 0.34 $\pm$ 0.068	& 0.47  $\pm$ 0.112	& 0.40	$\pm$ 0.122\\
E(V$-\rm R_{c}$) 	& 0.08 $\pm$ 0.068	&       		& 0.10	$\pm$ 0.031\\
E(V$-\rm I_{c}$) 	& 0.17 $\pm$ 0.073	& 0.27  $\pm$ 0.122	& 0.22	$\pm$ 0.069\\
E($\rm R_{c}-I_{c}$) 	& 0.09 $\pm$ 0.076	&       		& 0.12	$\pm$ 0.037\\

\hline
\end{tabular}
\label{table 3}
\end{table}

\begin{table}
\caption{Parameters of SN 2004S}
\begin{tabular}{@{}ll}\hline
Discovery Date&2004 February 3.54 UT\\
Host Galaxy&MCG-05-16-021\\
Galaxy Type&Morphological type S\\
RA (2000) & $06^{\rm h}45^{\rm m}43^{\rm s}.5$ $\pm$ $0^{''}.1$\\
Dec (2000) &$-31^{\circ}13^{'}52^{''}.5$ $\pm$ $0^{''}.1$ \\
Offset from the nucleus&$47^{'}$.2 W \& $2^{''}$.5 S\\
Spectrum&Type Ia\\
Radial velocity&2731 $\pm$ 36 \\
from galaxy redshift&\\
(km s$^{-1}$) & 2730 $\pm$ 42 (LEDA)\\
Radial velocity &2516 (LEDA)\\
corrected for LG  infall onto & \\
 Virgo (km s$^{-1}$) & \\
Expansion velocity&9300 km $\rm sec^{-1}$ \\
of the SN&\\
Distance modulus & $\mu$ = 32.94 mag  \\
($H_0$=65 km~s$^{-1}$~Mpc$^{-1}$) & \\
Time of B maximum (JD) &2453039.42 $\pm$ 0.64\\
(from template fitting)&\\
Magnitudes at $\rm t_{B_{max}}$&$\rm B({t_{B_{max}}})$ = $14.45 \pm 0.05$\\
&$\rm V({t_{B_{max}}})$ = $14.36 \pm 0.07$  \\
&$\rm R_{c}({{t_{B_{max}}}})$ = $14.39 \pm 0.07$ \\
&$\rm I_{c}({{t_{B_{max}}}})$ = $14.52 \pm 0.09$ \\
Absolute Magnitudes at $\rm t_{B_{max}}$ &$\rm M^{B}_{t_{B_{max}}}$ = $-19.05 \pm 0.23$ \\
&$\rm M^{V}_{t_{B_{max}}}$ = $-18.96 \pm 0.18$\\
&$\rm M^{R_{c}}_{t_{B_{max}}}$ = $-18.82  \pm 0.15$\\
&$\rm M^{I_{c}}_{t_{B_{max}}}$ = $-18.58  \pm 0.14$\\
Adopted Total Extinction&$A_{B}$ = $0.716 \pm 0.220$ \\
&$A_{V}$ = $0.542 \pm 0.167$\\
&$A_{R_{c}}$ = $0.439 \pm 0.135$\\
&$A_{I_{c}}$ = $0.320 \pm 0.098$\\
Magnitude of secondary & 15.020\\
 $\rm I_{c}$ maximum (from template)& \\
$\Delta m_{15}$ in $B$ from & 1.11\\ 
template&\\
$\Delta m_{15}$ in $B$ from&1.26 $\pm$ 0.061 \\
observations&\\ 
Decline rate per& B band 0.115 $\pm$ 0.029 \\
day $\sim$ 8 days after $\rm t_{B_{max}}$&V band 0.054 $\pm$ 0.014\\
&$\rm R_{c}$ band 0.055 $\pm$ 0.035\\
&$\rm I_{c}$ band 0.034 $\pm$ 0.048\\
\hline
\end{tabular}
\label{table 4}
\end{table}

We estimated the characteristic parameter $\Delta$${m_{15}}$, the number of 
magnitudes in B band by which the SN declines in the first 15 days after maximum. The 
fitted template has a $\Delta$${m_{15}}$ of 1.11. We also calculated 
$\Delta$${m_{15}}$ by taking the B band peak magnitude obtained by the template
 fit and the observed B band magnitude after $\sim$ 15 days of the B band peak.
 This gives a value of $\Delta$${m_{15}} = $ 1.26 $\pm$ 0.061. We estimated 
the average decline rate in all bands from our observations, using a time 
baseline of 10 days starting $\sim$ 8 days after the B band peak, when our 
observations began. These decline rates are listed in Table 4.
\begin{figure}
\includegraphics[width=84mm]{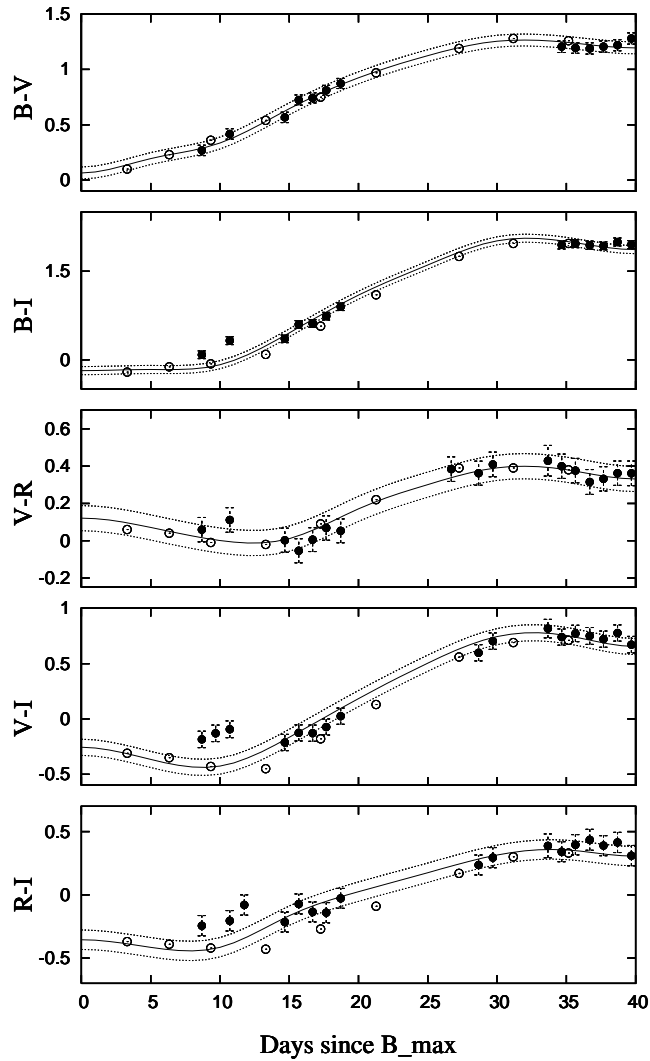}
\caption{Color curves of SN 2004S. The middle (solid) line in each panel
shows the extinction corrected colour curve bounded by errors 
(dashed lines) on both sides. Filled circles represent colours obtained
from our observations, and open circles are from observations reported
by others.}
\label{figure4}
\end{figure}

\section{Absolute luminosity and Bolometric light curve }

Assuming $\rm H_{0}$=65 km $\rm sec^{-1}$ $\rm Mpc^{-1}$ and the radial 
velocity of MCG-05-16-021 corrected for Local Group infall onto Virgo as 
$\rm v_{r}$ = 2516 km $\rm sec^{-1}$ as listed in LEDA 
(http://leda.univ-lyon1.fr/), we find a distance modulus of 32.94 mag.  
 The total extinction estimated using the intrinsic 
colour curves of \citet{Nobili} are mentioned in table 3 (column 2). 
From these, the absolute magnitudes estimated at the time of $\rm B({t_{B_{max}}})$ 
in different bands are $\rm M^{t_{B_{max}}}_{B}$ = $-19.05$ $\pm$ $0.23$, 
$\rm M^{t_{B_{max}}}_{V}$ = $-18.96$ $\pm$ $0.18$, 
$\rm M^{t_{B_{max}}}_{R_{c}}$ = $-18.82$ $\pm$ $0.15$ and 
$\rm M^{t_{B_{max}}}_{I_{c}}$ = $-18.58$ $\pm$ $0.14$. 
\citet{Altavilla} suggests another method for estimating absolute magnitude
using a relation between $\rm M^{max}$ and $\Delta$${m_{15}}$. Adopting the 
values of linear fit coefficients of $-19.61$ $\pm$ $0.04$ and $1.10$ $\pm$ $0.15$
 as given by \citet{Altavilla}, we obtain 
$\rm M^{B}_{max}$ = $-19.43$ $\pm$ $0.08$. The $\rm M^{max}_{B}$ values 
obtained by the above two methods are in good agreement with each other.

Since most of the flux from an SN Ia emerges in optical bands during the 
first few weeks \citep{Suntzeff1}, the integrated flux in 
UBV$\rm R_{c}$$\rm I_{c}$ bands provides a meaningful estimate of the 
bolometric luminosity, which is directly related to the amount of radioactive 
nickel synthesized and ejected in the explosion. Supplementing our data in  
BV$\rm R_{c}$$\rm I_{c}$ bands and with U band observations reported by  
\citet{Krisciunas2}, we construct a bolometric light curve using 
de-reddened magnitudes and the estimated distance modulus till $\sim$ 40 
days after $\rm t_{{B}_{max}}$. The first U band observation was 3.105 days
after $\rm t_{{B}_{max}}$. To estimate the contribution of U band at peak we 
assumed that the (U-B) colour remains constant from the peak to 3.105 days. 
 \begin{figure}
\includegraphics[width=84mm,height=84mm]{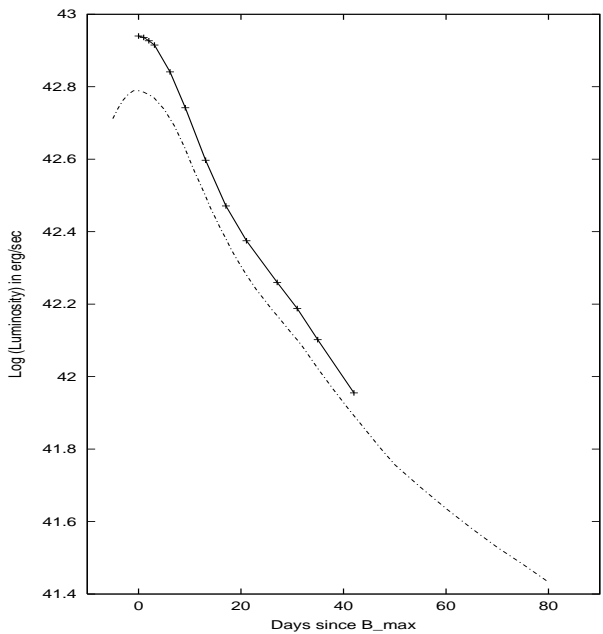}
\caption{Bolometric light curve for SN 2004S. The solid line shows the
Bolometric light curve constructed from UBV$\rm R_{c}$$\rm I_{c}$ bands and the
dash-dotted line shows the bolometric light curve derived from BVI fits of
\citet{Hamuy2}.}
\label{figure5}
\end{figure}
The magnitudes obtained were converted to flux using calibrations by 
\citet{Fukugita}. 
The contribution from the U band at $\rm t_{{B}_{max}}$ is $\sim$ 18.24 $\%$ and 
that from the $\rm I_{c}$ band is $\sim$ 11.12$\%$. We have not 
accounted for the contribution from JHK bands. In Figure 5 we show the UVOIR 
bolometric light curve as a solid line from 0 to 40 days after 
$\rm t_{{B}_{max}}$. The dash-dotted line in Figure 5 shows the contribution 
derived from the BVI bands alone, as obtained from fitted templates from 
$-5$ to $80$ days with reference to $\rm t_{{B}_{max}}$. We derive a peak bolometric
 luminosity of L = 8.715 x 10$^{42}$ erg $\rm sec^{-1}$. The bolometric peak is 
coincident in time with the B band peak within errors. We use the peak 
bolometric luminosity to derive $\rm ^{56}Ni$ mass using the method outlined 
by \citet{Branch2}. It is understood that radioactivity powers the light curve 
and near the time of maximum light most of the energy released by the 
radioactivity is still being trapped and thermalized. The peak radiated 
luminosity is expected to be comparable to the instantaneous rate of energy 
release by radioactivity \citep{Branch2}. The peak luminosity can be expressed 
as \\
\begin{center}
L = $\alpha \rm R(t_{R})M_{Ni}$
\end{center}
where $\alpha$ is a model dependent parameter expected to be of order unity, 
R is the radioactivity luminosity per unit nickel mass, evaluated at the time 
of maximum light $\rm t_{R}$ (the rise time) and $\rm M_{Ni}$ is the synthesized 
mass of $\rm ^{56}Ni$. The B band rise time ($\rm t_{R}$) is related to the 
post peak B band decline rate ($\beta$) in mag per 100 days as 
$\rm t_{R}$=13+0.7$\beta$. In the case of SN 2004S $\beta$ = 8.45,
which gives a rise time of 18.915 days. For this value of $\rm t_{R}$, 
$\rm R(t_{R})$ works out to be 
2.108 x 10$^{43}$ erg $\rm sec^{-1}$$M^{-1}_{\odot}$. 
The peak bolometric luminosity determined above then yields  
$\rm M_{Ni}$ = 0.41 $M_{\odot}$ for an assumed $\alpha$ = 1.

\citet{Contardo} have calculated the bolometric luminosity and the ejected 
nickel mass for several SN Ia from UBVRI bolometric peak fluxes. 
In Table 5 we compare $\rm M^{B}_{max}$, $\Delta$$m_{15}$, the peak bolometric 
luminosity and derived $\rm M_{Ni}$ for SN 2004S with corresponding values 
for the sample of \citet{Contardo} and two other recent SNe Ia: 1998bu 
\citep{Hernandez,Leibundgut3} and 1999aw \citep{Strolger}. We find that 
SN 2004S represents a mid range value for $\rm M_{Ni}$, similar to SN 1992A, 
SN 1992bo and SN 1994D. The subluminous event SN 1991bg is a fast decliner 
having a smaller value of Nickel mass ejected. SN 1991T, a peculiar and 
intrinsically bright supernova, has the largest value of derived $\rm M_{Ni}$ 
in the table.
More recently, observations of cepheids by HST has revised the distance
to NGC 4527, the host galaxy of SN 1991T \citep{Gibson}.  Also, possible
JHK maxima of this supernova have been determined by \citet{Krisciunas4}.
These measurements indicate that SN 1991T was only slightly overluminous,
comparable to Type Ia SNe with similar values of $\Delta m_{15}$.
\citet{Candia} point out another peculiar case of SN 2000cx as
an underluminous event. \citet{Candia} have compared the bolometric light
curves of SN 2000cx with SN 1999ee and SN 2001el. All three SN have
similar $\Delta m_{15}$, 0.93, 0.94 and 1.13 respectively.
However, \citet{Candia} also note that the underluminous nature of
SN 2000cx requires further confirmation with a
better distance estimate to the host NGC 524.

\begin{table}
\caption{Absolute B magnitude at peak, Peak Luminosity, $\Delta$$m_{15}$ and 
$\rm ^{56}Ni$ masses of few SN Ia}
\noindent
\begin{tabular}{@{}ccccc}\hline
SN&$\rm M_{B}$&$\Delta$$m_{15}$&log $\rm L_{bol}$&$\rm M_{Ni}$\\
&(mag)&& (erg $\rm sec^{-1}$)& $M_{\odot}$\\
\hline
1989B&-19.37&1.20&43.06&0.57\\
1991T&-20.06&0.97&43.23&0.84\\
1991bg&-16.78&1.85&42.32&0.11\\
1992A&-18.80&1.33&42.88&0.39\\
1992bc&-19.72&0.87&43.22&0.84\\
1992bo&-18.89&1.73&42.91&0.41\\
1994D&-18.91&1.46&42.91&0.41\\
1994ae&-19.24&0.95&43.04&0.55\\
1995D&-19.66&0.98&43.19&0.77\\
1998bu&-19.35&1.09&43.18&0.77\\
1999aw&-19.45&0.81&43.18&0.76\\
{\bf 2004S}&{\bf -19.05}&{\bf 1.26}&{\bf 42.94}&{\bf 0.41}\\
\hline
\end{tabular}
\label{table 4}
\end{table}

\section{Comparison with SN 1992al}

As mentioned in section 3 the template of SN 1992al fits best with our data 
set. We therefore compare the observed and derived properties of these two
supernovae in Table 6. For SN 1992al de-reddened and K-corrected apparent
 magnitudes, colour, decline rate and absolute magnitudes are taken from 
\citet{Hamuy1}. The corresponding quantities for SN 2004S are from this work. 
As we see from the table, the intrinsic properties of these two supernovae 
are very similar, although SN 2004S may be marginally less luminous 
particularly in the $\rm I_{c}$ band. The above comparison reinforces the general 
conclusion that for SNe Ia with similar light curves the intrinsic luminosities
 tend to be very similar. 
\begin{table*}
\caption{Comparison of the properties of SN 2004S with SN 1992al. Decline rate, 
apparent magnitudes corrected for galactic extinction, Color and Absolute 
magnitudes at the time of B band maximum are compared for SN 1992al and 
SN 2004S. The numbers in parenthesis are the errors in the respective 
parameters.}
\noindent
\begin{tabular}{@{}ccccccccc}\hline
SN&$\Delta$$m_{15}$&$\rm B_{max}$&$\rm V_{max}$&$\rm I_{max}$&$\rm B_{max}$-$\rm
 V_{max}$&$\rm M^{max}_{B}$&$\rm M^{max}_{V}$&$\rm M^{max}_{I_{c}}$\\
\hline
1992al&1.11(0.05)&14.60(0.07)&14.65(0.06)&14.94(0.06)&-0.05(0.03)&-19.47(0.32)&
-19.42(0.31)&-19.13(0.31)\\
2004S&1.262(0.061)&14.04(0.05)&14.05(0.07)&14.34(0.09)&-0.01(0.08)&-19.05(0.23)
&-18.96(0.18)&-18.58(0.14)\\
\hline
\end{tabular}
\label{table 6}
\end{table*}
 
\section{Conclusions }
 
 We report photometric observations of SN 2004S which were carried out using 
the 1-m Sampurnanand Telescope at ARIES, Nainital during 2004 February 12 to
March 22. UBV$\rm R_{c}$$\rm I_{c}$ light curves have been studied by combining
 our data with data available in the literature. We estimate the peak 
magnitudes in different bands and time of B maximum using the template fitting
 method. The light curve parameter $\Delta$$m_{15}$ is estimated  
to be 1.26 $\pm$ 0.061 from our data. The light curve template of SN 1992al 
fits quite well to SN 2004S. We present the bolometric light curve which 
illustrates the decay of total luminosity of the supernova. The estimated peak 
luminosity 8.715 x 10$^{42}$ erg/sec yields a value of $\rm ^{56}Ni$ mass 
ejected to be 0.41 $M_{\odot}$. Comparing the derived mass of ejected 
$\rm ^{56}Ni$ in different SNe Ia including SN 2004S we notice that for a 
given nickel mass, there could be a significant dispersion in peak luminosity,
 as the envelope structure and hence the decline rate parameter 
$\Delta$$m_{15}$ could be different in different cases. SN 2004S can be 
placed as a mid-range decliner and the ejected mass of $\rm ^{56}Ni$ also 
has a mid-range value in this case.\\
 
\section{ Acknowledgment}
We wish to thank J. C. Pandey and S. B. Pandey for their help with the 
photometric data reduction. Illuminating discussions with Abhijit Saha
are gratefully acknowledged. One of the authors (KM) also wishes to thank 
L. Resmi for several useful discussions. APK and KM acknowledge Dept. of 
Science and Technology, India for financial support. We are thankful to 
an anonymous referee for helpful comments and suggestions.

\end{document}